\documentclass[twocolumn,showpacs,preprintnumbers,amsmath,amssymb]{revtex4}

\usepackage{graphicx}
\usepackage{dcolumn}
\usepackage{bm}
\usepackage{color}

\begin{document}


\title{Small world yields the most effective information spreading}

\author{Linyuan L\"u$^{1,2}$}
\author{Duan-Bing Chen$^{1}$}
\author{Tao Zhou$^{1}$}
\email{zhutou@ustc.edu} \affiliation{$^1$Web Sciences Center,
University of Electronic Science and Technology of China, 610054
Chengdu, People's Republic of China \\ $^2$Department of Physics,
University of Fribourg, Chemin du Mus\'{e}e 3, CH-1700 Fribourg,
Switzerland}

\begin{abstract}

Spreading dynamics of information and diseases are usually analyzed
by using a unified framework and analogous models. In this paper, we
propose a model to emphasize the essential difference between
information spreading and epidemic spreading, where the memory
effects, the social reinforcement and the non-redundancy of contacts
are taken into account. Under certain conditions, the information
spreads faster and broader in regular networks than in random
networks, which to some extent supports the recent experimental
observation of spreading in online society [D. Centola, Science {\bf
329}, 1194 (2010)]. At the same time, simulation result indicates
that the random networks tend to be favorable for effective
spreading when the network size increases. This challenges the
validity of the above-mentioned experiment for large-scale systems.
More significantly, we show that the spreading effectiveness can be
sharply enhanced by introducing a little randomness into the regular
structure, namely the small-world networks yield the most effective
information spreading. Our work provides insights to the
understanding of the role of local clustering in information
spreading.

\end{abstract}


\maketitle

\section{Introduction}

Understanding the dynamics of epidemic spreading is a long-term
challenge, and has attracted increasing attention recently. Firstly,
the fast development of data base technology and computational power
makes more data available and analysable to scientific community.
Secondly, many new objects of study come into the horizon of
epidemiologists, such as computer virus, opinions, rumors,
behaviors, innovations, fads, and so on. Lastly, in addition to the
compartment model and population dynamics~\cite{May1991}, novel
models and tools appeared recently inspired by the empirical
discoveries about network
topology~\cite{Pastor-Satorras2001,Barrat2008}, temporal
regularities of human
activities~\cite{Vazquez2007,Iriburren2009,Yang2011} and scaling
laws in human mobility~\cite{Wang2009,Balcan2009}.

In the simplest way, we can roughly divide the human-activated
spreading dynamics into two classes according to the disseminules:
one is the spreading of infectious diseases requiring physical
contacts, and the other is the spreading of information including
opinions, rumors and so on (Here we mainly consider the information
whose value and authenticity need judge and verification by
individuals, different from the information about jobs, discounts,
etc.). In the early stage, scientists tried to describe these two
classes by using a unified framework and analogous models (see,
e.g., Ref.~\cite{Liu2003,Moreno2004}), emphasizing their homology
yet overlooking their essential differences. Very recently,
scientists started to take serious consideration about the specific
features of information spreading~\cite{Hill2010,House2011}, as well
as the different mechanisms across different kinds of
information~\cite{Romero2011}. Dodds and Watts \cite{Dodds2004}
studied the effects of limited memory on contagion, yet did not
consider the social reinforcement. Some recent works indicate that
the social reinforcement plays important role in the propagation of
opinions, news, innovations and fads
\cite{Centola2007,Medo2009,Cimini2011,Wei2011,Krapivsky2011}.

\begin{figure}
\begin{center}
\includegraphics[width=8cm]{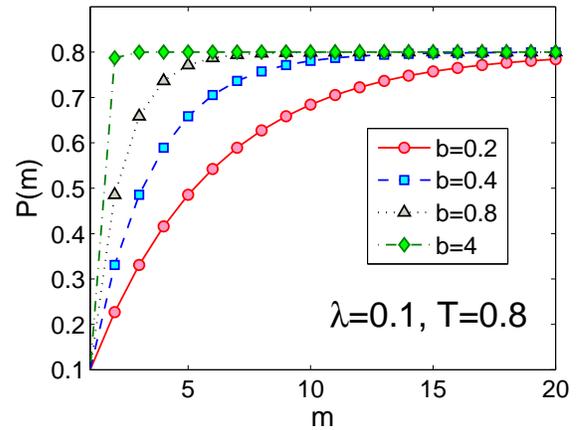}
\caption{(Color online) The approving probability as a function of
$m$. }\label{function}
\end{center}
\end{figure}

In this paper, we propose a variant of the
susceptible-infected-recovered (SIR) model for information
spreading, which takes into account three different spreading rules
from the standard SIR model: (i) memory effects, (ii) social
reinforcement, and (iii) non-redundancy of contacts. The main
contributions are twofold. Firstly, we show that when the spreading
rate $\lambda$ is smaller than a certain value $\lambda^*$, the
information spreads more effectively in regular networks than in
random networks, which to some extent supports the experiment
reported by Centola \cite{Centola2010}: behavior spreads faster and
can infects more people in a regular online social network than in a
random one (with no more than 200 people in the experiment). We
further show that as the increasing of the network size, the value
of $\lambda^*$ will decrease, which challenges the validity of
Centola's experiment \cite{Centola2010} for very large-scale
networks. Secondly, the effectiveness of information spreading can
be remarkably enhanced by introducing a little randomness into the
regular structure, namely the small-world networks \cite{Watts1998}
yield the most effective information spreading. This result is
complementary to the traditional understanding of epidemic spreading
on networks where the infectious diseases spread faster in random
networks than in small-world networks.

\begin{figure}
\begin{center}
\includegraphics[width=4.25cm]{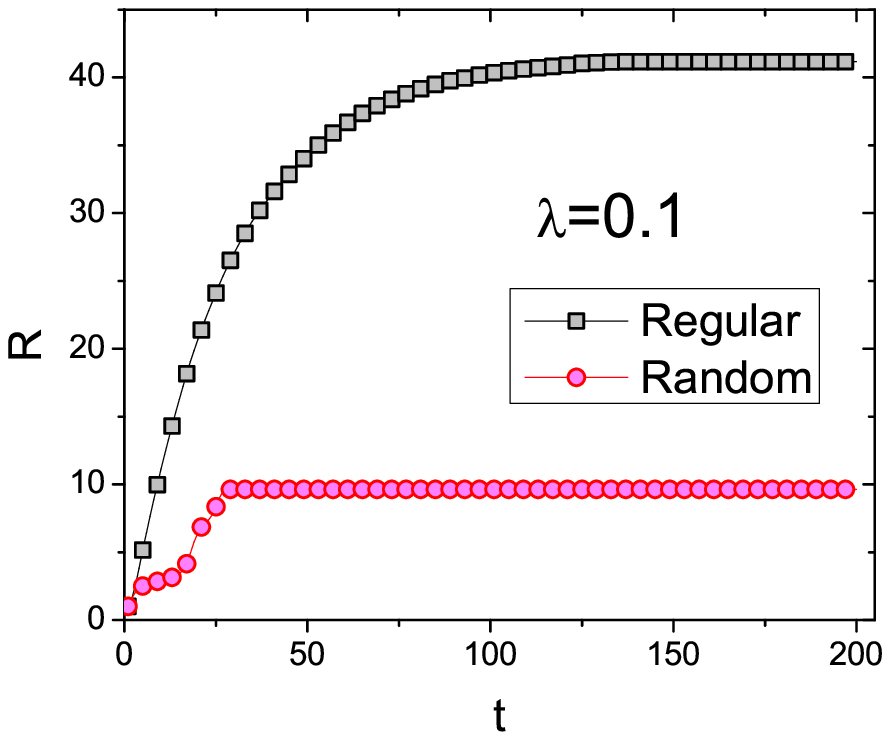}
\includegraphics[width=4.25cm]{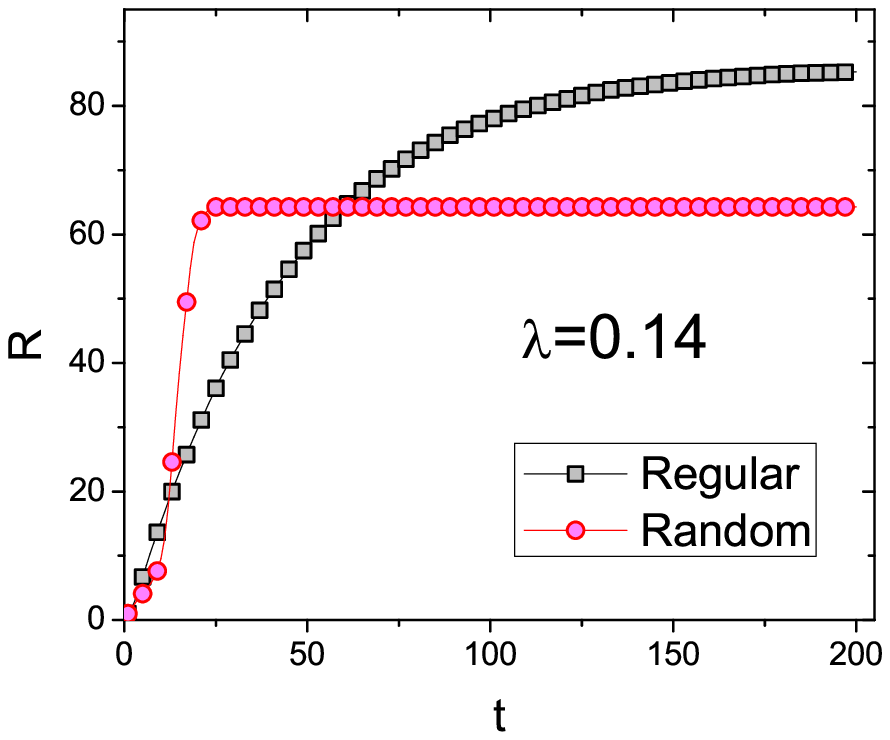}
\includegraphics[width=4.25cm]{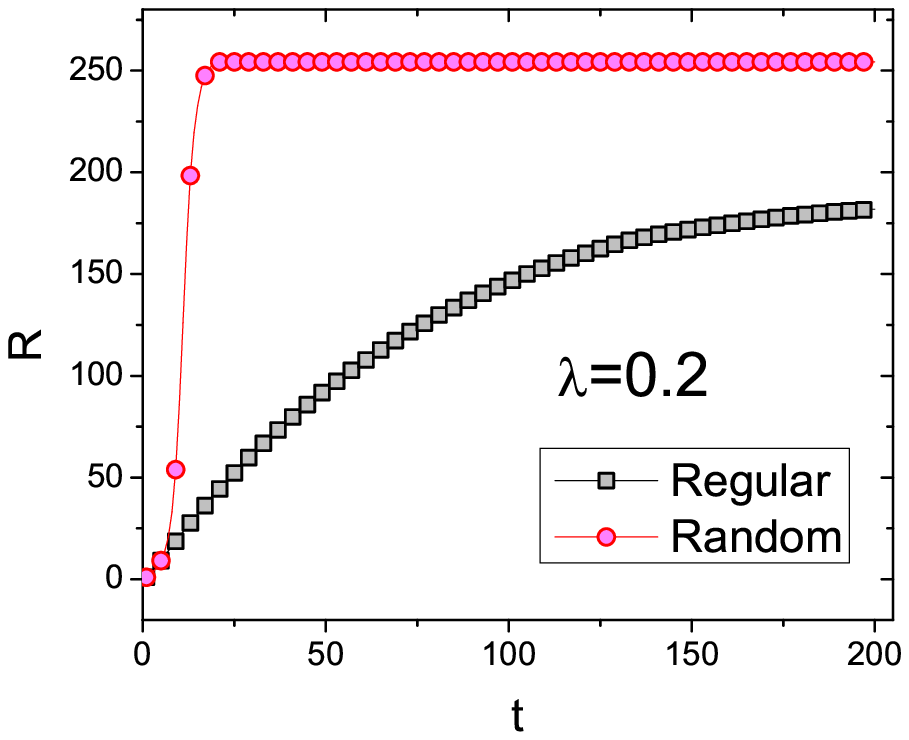}
\includegraphics[width=4.25cm]{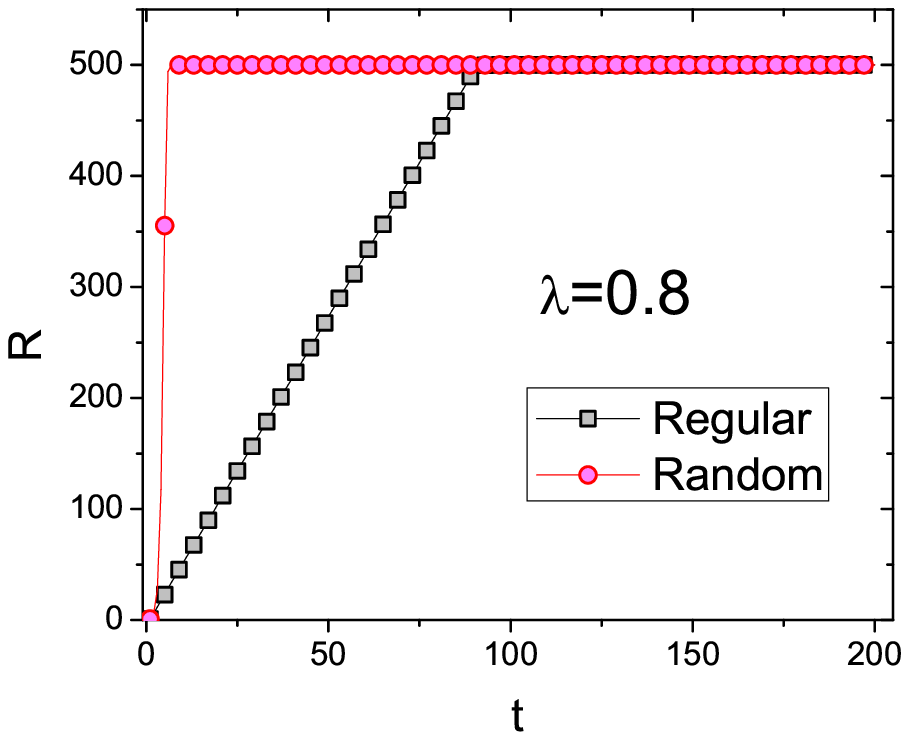}
\caption{(Color online) The number of approved nodes as a function
of time on regular network (black squares) and random network (red
circles). The parameters are $N=500$, $k=6$, $b=0.8$ and $T=1$. The
results are obtained by averaging over 500 independent
realizations.}\label{fig2}
\end{center}
\end{figure}

\section{Model}

Given a network with $N$ nodes and $E$ links representing the
individuals and their interactions, respectively. Hereinafter, for
convenience, we use the language of news spreading, but our model
can be applied to the spreading of many kinds of information like
rumors and opinions, not limited to news. At each time step, each
individual adopts one of four states: (i) \emph{Unknown}--the
individual has not yet heard the news, analogous to the susceptible
state of the SIR model. (ii) \emph{Known}--the individual is aware
of the news but not willing to transmit it, because she is
suspicious of the authenticity of the news. (iii)
\emph{Approved}--the individual approves the news and then transmits
it to all her neighbors. (iv) \emph{Exhausted}--after transmitting
the news, the individual will lose interest and never transmit this
news again, analogous to the recovered state in the SIR model.

At the beginning, one node is randomly chosen as the ``seed'' and
all others are in the unknown state. This seed node will transmit
the news to all her neighbors, and then become exhausted. Once an
individual (either in unknown or known state) receives a news, she
will judge whether it is true depending on the number of times he
has heard it---a news or a rumor is more likely to be approved if
being heard many times (a very recent model allows the infectivity
and/or susceptibility of hosts to be dependent on the number of
infected neighbors \cite{Perez-Reche2011}). The present rules imply
two features of information spreading, namely the memory effects and
social reinforcement, which are usually neglected in the standard
SIR model and its variants for rumor propagation.

\begin{figure}
\begin{center}
\includegraphics[width=8cm]{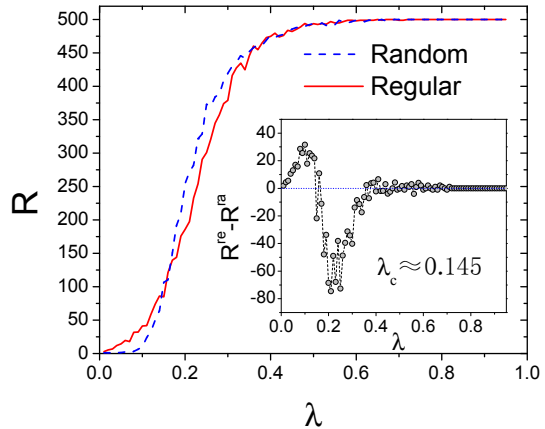}
\caption{(Color online) The dependence of the number of approved
nodes at the final state on the parameter $\lambda$ for regular (red
solid line) and random (blue dash line) networks. The parameters are
$N=500$, $k=6$, $b=0.8$ and $T=1$, as the same as those for
Fig.~\ref{fig2}. Inset shows the number of final approved nodes on
regular network $R^{\texttt{re}}$ minus that on random networks
$R^{\texttt{ra}}$, against $\lambda$. The results are obtained by
averaging over 500 independent realizations.}\label{fig3}
\end{center}
\end{figure}

In our model, we assume that for a given individual if she receives
the news at least once at the $t$th time step, and she has received
$m(t)$ times of this news until time $t$ ($m(t)$ is a cumulative
number), the probability she will approve it at time $t$ is
$P(m)=(\lambda-T)e^{-b(m-1)}+T$, where $\lambda=P(1)$ is the
approving probability for the first receival. $T\in (0,1]$ is the
upper bound of the probability indicating the maximal approving
probability. Here we do not consider the interest decay, and we
assume that the time scale of news spreading is much faster than our
memory decay. After approval, she will transmit the news to all her
neighbors in the next time step and then turn to be exhausted. If an
individual, either in unknown or known states, does not receive any
news in the $t$th time step, nothing will happen no matter how many
times this individual has received the news. The memory effects are
embodied by $m(t)$ which is a cumulative number instead of the
independent spreading rates for different contacts in the standard
SIR model. With the increasing of $m$, $P(m)$ will infinitely
approach to $T$ and the speed is determined by the parameter $b>0$
which reflects the social reinforcement effect. Figure
~\ref{function} shows the approving probability as a function of
$m$, given different $b$. Larger $b$ indicates a stronger social
reinforcement. For example, $P(2)=0.227$ when $b=0.2$, and it equals
0.486 for $b=0.8$. Since an individual who has transmitted the news
will immediately become exhausted, our model ensures that each link
is used at most once without any redundancy of contacts. The
spreading process comes to the end when no new individual approves
the news and spreads it.

\begin{figure}
\begin{center}
\includegraphics[width=8cm]{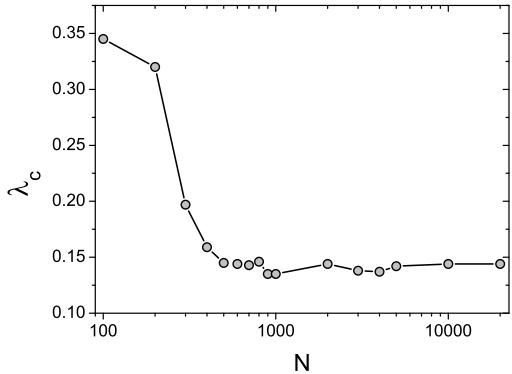}
\caption{The dependence of $\lambda_c$ on the network size $N$. The
parameters are $k=6$, $b=0.8$ and $T=1$. The results are obtained by
averaging over 500 independent realizations.}\label{fig4}
\end{center}
\end{figure}

We perform our model on three kinds of networks with identical node
degree $k$. (i) \emph{Regular networks}.--A regular network is a one
dimensional ordered network with periodic boundary conditions, where
each node is connected to its $k$ nearest neighbors, namely to the
$k/2$ nearest neighbors clockwise and counterclockwise
\cite{Watts1998}. Notice that, in the literature of graph theory,
the term ``regular networks" usually stands for networks whose nodes
are of the same degree, and thus the following homogeneous
small-world networks are also regular. In this article, we follow
the literature of complex networks and use the term ``regular
networks" to represent networks with ordered structure. (ii)
\emph{Homogeneous small-world networks}.--The homogenous small-world
network is constructed by randomly reshuffling links of a regular
network, while keeping the degree of each node unchanged
\cite{Santos2005}. According to the link exchanging method
\cite{Maslov2002}, at each time step, we randomly select a pair of
edges A-B and C-D. These two edges are then rewired to be A-D and
B-C. To prevent the multiple edges connecting the same pair of
nodes, if A-D or B-C already exists in the network this step is
aborted and a new pair of edges is randomly selected. We implement
$pE$ steps, where $p$ indicates the randomness of the network. (iii)
\emph{Random networks}.--Repeating the above rewiring operations
many times leads to a homogenous random network. Theoretically
speaking, a homogenous random network is obtained only for
$p\rightarrow \infty$, we here consider $p\in [0,10]$ and when
$p>1$, the topological statistics are very close to the ones of
random networks. In all simulations, the node degree is set to be
$k=6$, and we have carefully checked that the results are not
sensitive to the node degree unless $k$ is very large or very small.

\section{Results}

\begin{figure}
\begin{center}
\includegraphics[width=9cm]{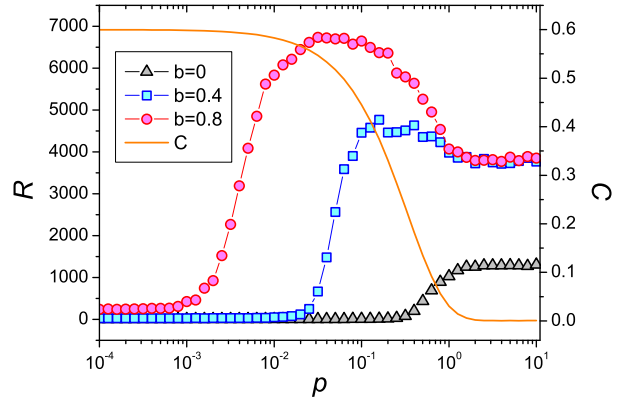}
\caption{(Color online) The number of final approved nodes against
the randomness $p$ given $b=0$ (triangles), $b=0.4$ (squares) and
$b=0.8$ (circles). Other parameters are $N=10^4$, $k=6$,
$\lambda=0.2$ and $T=1$. The results are obtained by averaging over
10000 independent realizations. The clustering coefficient $C$, as a
monotonic function of $p$, is also displayed.} \label{fig5}
\end{center}
\end{figure}

Denote by $R$ the number of approved nodes of the news. Larger $R$
at the final state indicates a broader spreading. We firstly compare
the spreading processes on regular and random networks. Figure
\ref{fig2} reports four typical examples with different $\lambda$
and fixed $b=0.8$. Surprisingly, for small $\lambda$ (e.g.,
Fig.~\ref{fig2}(a)), the spreading on regular networks is faster and
broader than on random networks. These results are in accordance
with the online social experiment of Centola \cite{Centola2010}, yet
against the traditional understanding of network spreading
\cite{Zhou2006}. With the increasing of $\lambda$, the random
networks will be favorable for faster and broader spreading.
Figure~\ref{fig3} shows the dependence of the number of approved
nodes at the final state on the parameter $\lambda$. There is a
crossing point at about $\lambda_c\approx 0.145$, after which $R$ of
random networks exceeds that of regular networks. The inset shows
the difference between numbers of final approved nodes on regular
and random networks, namely $R^{\texttt{re}}-R^{\texttt{ra}}$
against $\lambda$. With very large $\lambda$, almost every node will
run into the approved state, and thus $R$ is not sensitive to the
network structure, but the spread on random networks is still faster
than on regular networks (see, for example, Fig.~\ref{fig2}(d)).

Figure 4 displays the crossing point $\lambda_c$ as a function of
the network size $N$. When $N$ is small, $\lambda_c$ decreases
sharply with the increasing of $N$, while when $N$ gets larger
$\lambda_c$ becomes insensitive to $N$. As a whole, $\lambda_c$
shows a non-increasing behavior versus $N$. Notice that, the
phenomenon that spreading on regular networks is faster and broader
than on random networks is more remarkable and easier to be observed
if $\lambda_c$ is large. Therefore, our result about $\lambda_c(N)$
indicates that for large-scale systems, Centola's experimental
results may be not hold or will be weaken to some extent.

\begin{figure}
\begin{center}
\includegraphics[width=8cm]{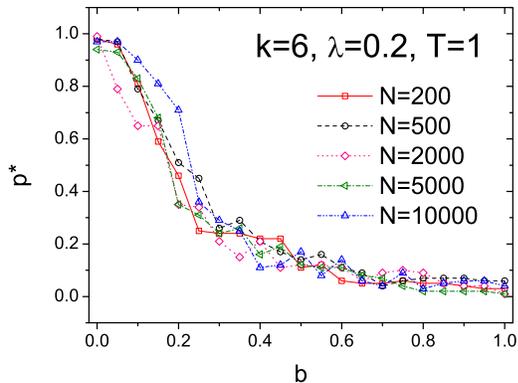}
\caption{(Color online) The dependence of optimal randomness $p^*$
on the strength of social reinforcement $b$ given different $N$. The
results are obtained by averaging over 500 independent
realizations.}\label{fig6}
\end{center}
\end{figure}

In previous study on SIR model, it was pointed out that the number
of recovered nodes at the end of evolution increases with the
increasing of randomness $p$ in small-world networks
\cite{Zanette2001R}. In contrast, our simulations show that the
number of approved nodes in the final state does not monotonously
increase with the increasing of $p$, instead, an optimal randomness
$p^*$ exists subject to the highest $R$. Figure~\ref{fig5} shows the
dependence of the number of final approved nodes on the randomness
$p$ given $b=0$ (triangles), $b=0.4$ (squares) and $b=0.8$
(circles). With strong social reinforcement, even a very small
randomness can bring a remarkable improvement of the number of final
approved nodes, $R$. Take the case $b=0.8$ for example, on the
regular networks (i.e., $p=0$), $R$ is 205, while by introducing a
tiny randomness $p=0.02$, this number will suddenly increase to
6593, which is also higher than the random networks (i.e., $p=1$,
$R=4049$). We also plot the clustering coefficient $C$ as a function
of $p$ in figure 5. As expected, $C$ decreases as the increasing of
$p$. The results indicate that the local clustering can to some
extent enhance the approving rate of information, which refine the
completely negative valuation of clustering coefficient in epidemic
spreading \cite{Eguiluz2002,Petermann2004,Zhou2005}.

The dependence of optimal randomness $p^*$ on the strength of social
reinforcement $b$ given different $N$ are shown in Fig.~\ref{fig6},
where one can observe that the stronger social reinforcement (i.e.,
larger $b$) results in a smaller $p^*$. In the presence of weak
social reinforcement (i.e., small $b$), our result ($p^*$ is close
to 1) is analogous to the well-known one
\cite{Zhou2006,Zanette2001R} that the speed and range of spreading
obey the relation ``Random $>$ Small-World $>$ Regular". In
contrast, the small-world networks yield the most effective
spreading when the social reinforcement plays a considerable role
(i.e., large $b$).

\begin{figure}
\begin{center}
\includegraphics[width=8cm]{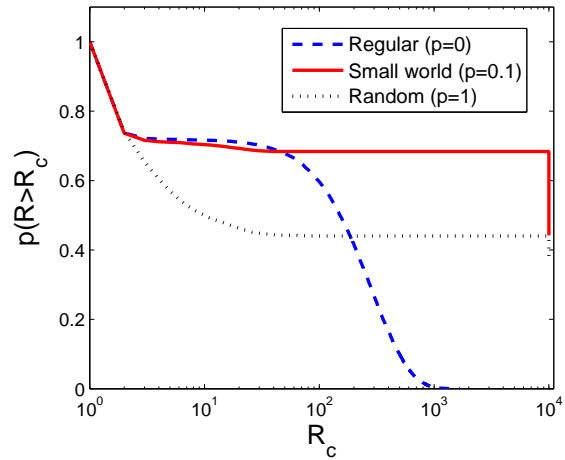}
\caption{(Color online) The cumulative probability that in a
realization, the information reaches more than $R_c$ individuals for
regular, random and small-world networks. The parameters are
$\lambda=0.2$, $k=6$, $T=1$ and $N=10000$. These distributions are
obtained from 10000 realizations.}\label{fig7}
\end{center}
\end{figure}

To further investigate the advantages of small-world networks for
information spreading, we calculate the complementary cumulative
distribution $p(R>R_c)$, namely the probability that in a
realization the information has reached more than $R_c$ individuals.
As shown in figure 7, comparing with random networks, the advantages
of small-world networks are twofold. On one hand, it has higher
probability to spread out (see the region when $R_c$ is small). For
example, in small-world networks $p(R>10)=0.703$, while for random
networks, this number is only 0.460. If the information can spread
out, like an epidemic for a disease, in both two kinds of networks
it can reach the majority of population. In contrast, comparing with
regular networks, information in small-world networks can spread
wide. According to figure 7, the maximum $R$ in regular networks is
only 1680 while in small-world networks it can reach 9900
individuals with probability 0.684.

\section{Conclusion and Discussion}

Thanks to the fast development of data base technology and
computational power, the detailed analysis about information
spreading in large-scale online systems become feasible nowadays. In
our opinion, the similarity between information spreading and
epidemic spreading are over emphasized in the previous studies (see,
for example, the models summarized in the review article
\cite{Castellano2009}), and currently we should turn to the other
side of the matter: revealing the essential difference between them.
The significant difference may include: (i) \emph{Time decaying
effects}.-- An infectious disease can exist more than thousands of
years in human society and still keep active, but no one is willing
to spread a news one year ago. Actually, our attention on
information decays very fast \cite{Wu2007}, and thus when we model
the information spreading, especially if it involves multiple
information competing for attention, we have to consider the time
decaying effects. (ii) \emph{Tie strength}.-- It is well known that
in social networks, ties with different strengths play different
roles in maintaining the network connectivity \cite{Onnela2007},
information filtering \cite{Lu2010}, information spreading
\cite{Miritello2011}, and so on. We guess the weak ties provide
faster paths for information spreading while the strong ties provide
trust paths (i.e., with high infectivity). However, this point is
still not clear till far. (iii) \emph{Information Content}.--
Information with different contents may have far different spreading
paths, and even with the same content, different expressions may
lead to far different performances. Some of them are born with
fashionable features while others are doomed to be kept from known.
Whether these two kinds of information are only different
quantitatively or they follow qualitatively different dynamic
patterns are still under investigation \cite{Crane2008}. (iv)
\emph{Role of spreaders}.-- Recent analysis on Twitter show that
different kinds of spreaders, such as media, celebrities, bloggers
and formal organizations, play remarkably different roles in network
construction and information spreading \cite{Wu2011}, which may
result in different spreading pathes and outbreaking mechanisms from
epidemic spreading. (v) \emph{Memory effects}.-- Previous contacts
could impact the information spreading in current time
\cite{Dodds2004}. Such memory effects can be direct since an agent
may tend to be interested in our disgusted with objects heard many
times, and/or indirect since previous contacts could change the tie
strength that further impact the current interactions. (vi)
\emph{Social reinforcement}.-- If more than one neighbor approved
the information and transferred it to you, you are of high
probability to approve it. Generally speaking, is an agent receives
twice an information item recommended from her neighbors, the
approval probability should be much larger than the twice of the
approval probability with a single recommending. (vii)
\emph{Non-redundancy of contacts}.-- People usually do not transfer
an information item more than once to the same guy, which is far
different from the sexually transmitted diseases. To name just a
few.

In this paper, we propose a simple model for information spreading
in social networks that considers the memory effects, the social
reinforcement and the non-redundancy of contacts. Under certain
conditions, the information spreads faster and broader in regular
networks than in random networks, which to some extent supports the
Centola's experiment \cite{Centola2010}. At the same time, we show
that the random networks tend to be favorable for effective
spreading when the network size increases, which challenges the
validity of the Centola's experiment for large-scale systems.
Furthermore, simulation results suggest that by introducing a little
randomness into regular structure, the small-world networks yield
the most effective information spreading. Although this simple model
cannot take into account all the above-mentioned features in
information spreading, it largely refines our understanding about
spreading dynamics. For example, traditional spreading models on
complex networks show that the diseases spread faster and broader in
random networks than small-world networks
\cite{Zhou2006,Zanette2001R}, yet our results suggest that the small
world may be the best structure for effective spreading under the
consideration of social reinforcement. Indeed, information in
small-world networks has much higher probability to spread out than
in random networks, and can spread much broader than in regular
networks. In addition, the local clustering is well-known to play a
negative role in spreading
\cite{Eguiluz2002,Petermann2004,Zhou2005}, while our model indicates
that local clustering are very helpful in facilitating the
acceptance/approval of the information for individuals and thus can
to some extent fasten the spreading.

\begin{acknowledgments}
We acknowledge Xiao-Ke Xu for valuable discussion. This work is
supported by the National Natural Science Foundation of China under
Grant No. 90924011, the Fundamental Research Funds for the Central
Universities, and the Swiss National Science Foundation under Grant
No. 200020-132253.
\end{acknowledgments}

\end{document}